\documentclass[aps,pre,twocolumn,groupedaddress,showpacs]{revtex4}
\usepackage{amssymb}
\usepackage{graphicx,color}
\definecolor{r}{rgb}{1,0,0}
\definecolor{g}{rgb}{0,1,0}
\definecolor{b}{rgb}{0,0,1}

\begin{document}


\title{Dynamical heterogeneity in soft particle suspensions under shear}


\author{K. N. Nordstrom$^1$, J. P. Gollub$^{1,2}$, and D. J. Durian$^1$}
\affiliation{$^{1}$Department of Physics and Astronomy, University of Pennsylvania, Philadelphia, PA
19104-6396, USA} \affiliation{$^{2}$Department of Physics and Astronomy, Haverford College,
Haverford, PA 19041-1392, USA}


\date{\today}

\begin{abstract}
We present experimental measurements of dynamical heterogeneities in a dense system of microgel spheres, sheared at different rates and at different packing fractions in a microfluidic channel, and visualized with high speed digital video microscopy.  A four-point dynamic susceptibility is deduced from video correlations, and is found to exhibit a peak that grows in height and shifts to longer times as the jamming transition is approached from two different directions.  In particular, the time for particle-size root-mean square relative displacements is found to scale as $\tau^* \sim (\dot \gamma \Delta \phi^4)^{-1}$ where $\dot\gamma$ is the strain rate and $\Delta\phi=|\phi-\phi_c|$ is the distance from the random close packing volume fraction.  The typical number of particles in a dynamical heterogeneity is deduced from the susceptibility peak height and found to scale as $n^* \sim (\dot \gamma \Delta \phi^4)^{-0.3}$.  Exponent uncertainties are less than ten percent.  We emphasize that the same power-law behavior is found at packing fractions above and below $\phi_c$.  Thus, our results considerably extend a previous observation of $n^* \sim \dot\gamma^{-0.3}$ for granular heap flow at fixed packing below $\phi_c$. Furthermore, the implied result $n^*\sim (\tau^*)^{0.3}$ compares well with expectation from mode-coupling theory and with prior observations for driven granular systems.
\end{abstract}

\pacs{64.70.pv, 83.80.Kn, 05.20.Jj}

%
%

\maketitle







\section{Introduction}

Disordered materials of all kinds are considered to be ``jammed'' if the relaxation time grows longer than the observation window, so that that the constituent particles appear locked into a fixed configuration of nearest neighbors \cite{LiuNagelBOOK, ohern03, LiuNagelARCMP10}.  For example, supercooled liquids can become jammed by lowering the temperature; hard sphere colloidal particles can become jammed by increasing the density; macroscopic glass beads can become jammed by lowering a driving force below some threshold.  No matter what the material or set of control parameters, as jamming is approached it has long been assumed that the growing relaxation time is accompanied by increasing co-operativity in particle motion \cite{AdamGibbs}.  The closer to jamming, the larger the number of neighbors that must cooperate in order to rearrange and the less frequently this happens.

It is now widely accepted that the rearrangement dynamics are not continuous near jamming, but rather are spatially and temporally heterogeneous \cite{LucaDHbook}.  Intermittent string-like swirls of rearranging particles come and go in a background of less mobile particles.  The four-point dynamical susceptibility $\chi_4(\tau)$ is a powerful tool for characterizing such dynamical heterogeneities \cite{lacevic03, ToninelliPRE05}.  This function exhibits a peak at a characteristic relaxation time, $\tau^*$, and the peak height $\chi_4^*$ can be related by a counting argument to the number $n^*$ of particles in the fast-moving rearranging regions \cite{abate07}.  One of the central questions today, then, is the quantitative relationship between the respective growth of $\tau^*$ and of $\chi_4^*$ on approach to jamming.  Expectations for various models are reviewed in Ref.~\cite{ToninelliPRE05}.  For example a logarithmic connection is expected for ``collectively-rearranging region'' scenarios.  A power-law connection $\chi_4^*\propto(\tau^*)^\lambda$ is predicted by mode-coupling theory, where $\lambda$ is the reciprocal of the mode-coupling exponent, $\gamma$; Ref.~\cite{ToninelliPRE05} particularly notes the values $\lambda=0.37$ \cite{KobPRL94} and $\lambda=0.40$ \cite{WhitelamPRL04}.  A power-law connection with $\lambda=1$ is expected for freely-diffusing defects.  And more recently a value $\lambda=1/2$ was reported for a kinetically constrained model jamming model \cite{YairEPL10}.


For colloidal hard spheres this issue was recently explored in Ref.~\cite{BrambillaPRL09}, which improves upon pioneering observations \cite{SillescuL98, marcus99, kegel00, weeks00} by covering an unprecedented density range near jamming such that the structural relaxation time increased by seven orders of magnitude.  The data show that $\tau^*$ grows faster than a power law, and $n^*$ grows slower than a power law, in $1/(\phi_c-\phi)$ as $\phi$ approaches $\phi_c$ from below.  The critical packing fraction $\phi_c$ is close to, but possibly distinct from, random close packing.  Irrespective of the value, the conclusion is that $n^*$ grows logarithmically with $\tau^*$.  For other colloidal systems it is not yet known whether this relationship depends on the nature of the particle interactions, or whether it changes when the control parameter is temperature or driving rather than just density.  In this paper, we report on dynamical heterogeneities for dense suspensions of soft Hertzian colloidal particles.  In particular we measure the time and size scales, and determine how they grow as jamming is approached both by bringing the density toward $\phi_c$, from either side, and also by lowering the strain rate.  In addition to establishing the dependence of $n^*$ and $\tau^*$ on these two control parameters, we also show that the size and time scales are related by a power law.  This result contrasts with Ref.~\cite{BrambillaPRL09}, but compares well with observations for macroscopic hard spherical grains, where the control parameters are density and fluidizing air speed \cite{AbatePRL08} or  strain rate and depth into a flowing heap \cite{katsuragi10}.  Here, as in Ref.~\cite{BrambillaPRL09} and Ref.~\cite{katsuragi10}, the dynamic range in relaxation time is more than seven orders of magnitude.


\section{Experimental Details}

\begin{figure}
\includegraphics[width=2.8in]{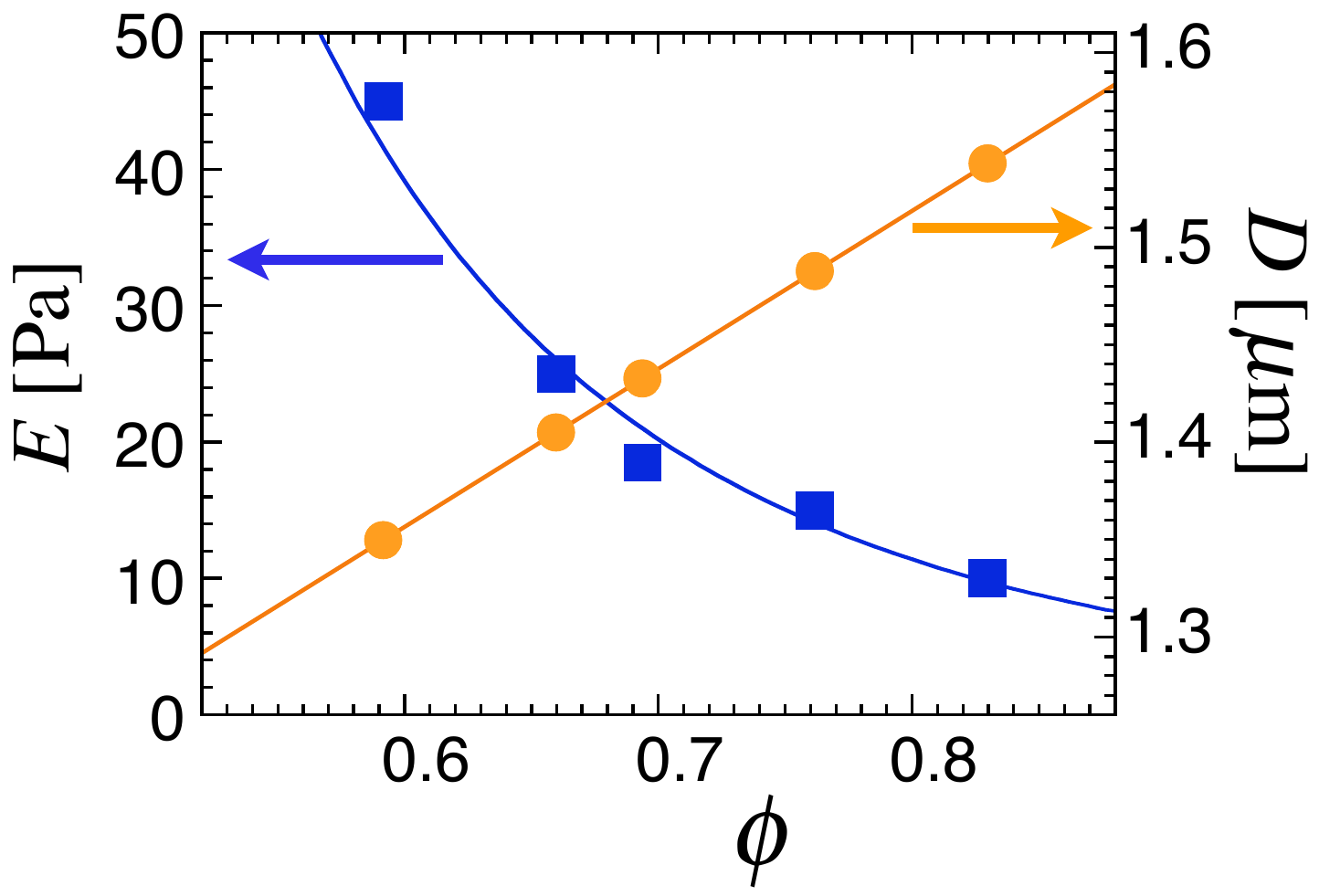}
\caption{(Color online) Properties of NIPA microgels as a function of packing fraction, as controlled by temperature:  (left) the elastic modulus measured by centrifugal compression, and (right) the particle diameter measured by dynamic light scattering \protect{\cite{CentComp}}.}
\label{nipaproperties}
\end{figure}

\begin{figure}
\includegraphics[width=2.in]{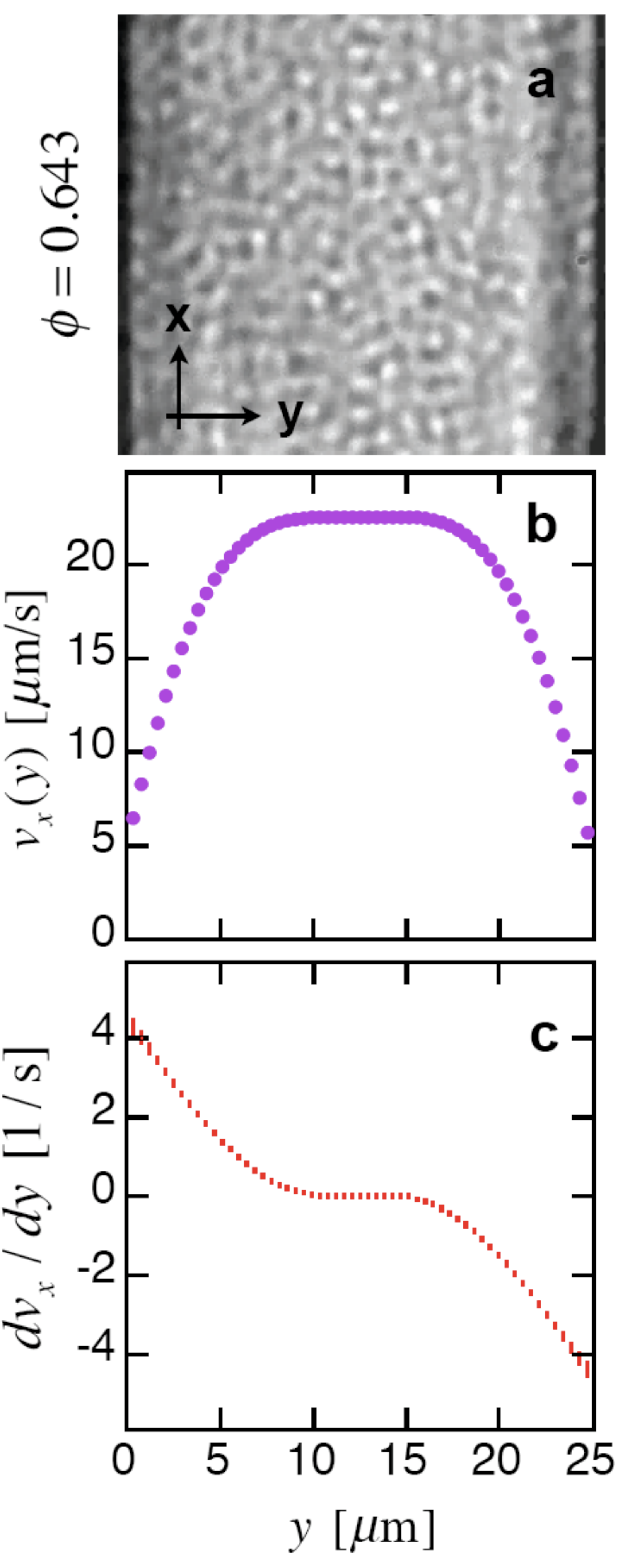}
\caption{(Color online) (a) An image taken from video data of particles.  (b) The velocity profile across the channel, extracted using PIV, and  (c) the strain rate as a function of position.} 
\label{schematic}
\end{figure}

The system we study is a dense aqueous suspension of thermoresponsive N-isopropylacrylamide (NIPA) microgel beads \cite{SaundersACIS99, PeltonACIS00}, synthesized with the Yodh group at Penn \cite{alsayed05, zhang09, yunker09, yunker10, YodhPRL10}.  Dynamical heterogeneities for unsheared suspensions of such particles have been reported previously in Refs.~\cite{sessoms09, yunker09, ColinSM10}, both below random close packing as well as above -- where aging effects are important.  Here two different size particles are used, primarily about 1~$\mu$m but also about 0.6~$\mu$m in diameter; for the former, the number density is $0.455/\mu{\rm m}^3$ and the viscosity of the suspending water is $\eta_0=0.01$~g/(cm-s).    Fig.~\ref{nipaproperties} shows the diameter and Young elastic modulus $E$ for the larger particles, obtained previously from dynamic light scattering and centrifugal compression \cite{CentComp} respectively.  Note that decreasing the temperature causes the particles to swell with water and become softer.  The applied pressure needed to squeeze water from the gel is very large compared to the elastic modulus \cite{CentComp}; therefore, the particles deform without deswelling and can be compressed to a known volume fraction $\phi$ above random close packing $\phi_c=0.635$, simply by lowering temperature.

Previously we studied the shear rheology of these suspensions by a custom microfluidic technique, in which the velocity profile is measured at the mid-height of a tall channel for various packing fractions and for various pressure-controlled flows \cite{Nordstrom2010}.  The channel is 25~$\mu$m wide, 100~$\mu$m tall, and $L=2$~cm long, fabricated of PDMS by soft lithography and bonded to a glass microscope slide.  The suspension is forced through the channel using pressurized air and inlet/outlet tubing of sufficient diameter that the imposed pressure drop $\Delta P$ occurs only along the length $L$ of the channel and can be related to the local shear stress in the suspension as $\sigma(y)=\Delta P y/L$.  The local shear strain rate is found by numerical differentiation of the velocity profile, $\dot\gamma(y)={\rm d}v_x(y)/{\rm d}y$. For this, we collect video data with a Phantom CMOS camera (1-10,000~fps) connected to a Zeiss Axiovert 200 microscope with $100\times$ objective.  An objective-cooling collar (Bioptechs) and cooling plate above the sample are controlled to about 0.1~C in order to vary the particle volume fraction.  An example video frame in Fig.~\ref{schematic}a displays bead-scale intensity variations, so that Particle Image Velocimetry (PIV) may be implemented with custom LabVIEW code. In practice, we break the image into strips 7 pixels wide and determine the velocity and strain rate in these regions. Example velocity and strain rate profile data are shown in Figs.~\ref{schematic}b-c.  In this figure the packing fraction is $\phi=0.643$, which is slightly above random close packing $\phi_c=0.635$; therefore, the flow is somewhat plug-like and exhibits wall slip.  We found that the resulting stress vs strain rate shear rheology could be collapsed onto two branches, by Olsson-Teitel \cite{olsson07} scaling with powers of the distance $\Delta \phi = |\phi - \phi_c|$ to jamming; the resulting exponents can be understood in terms of particle interactions \cite{Tighe2010}.

The full data set from Ref.~\cite{Nordstrom2010} consists of video plus stress and strain rate profiles for packing fractions varied discretely between 0.5 and 0.7, and for strain rates varied continuously from $10^{-3}$ to 100~s$^{-1}$.  Since the systems are always under shear, steady state conditions are attained where there are no aging effects.  In the following sections, we analyze the same data for dynamical heterogeneities.

 
\section{Heterogeneity Analysis}

\begin{figure}
\begin{center}
\includegraphics[width=2.8in]{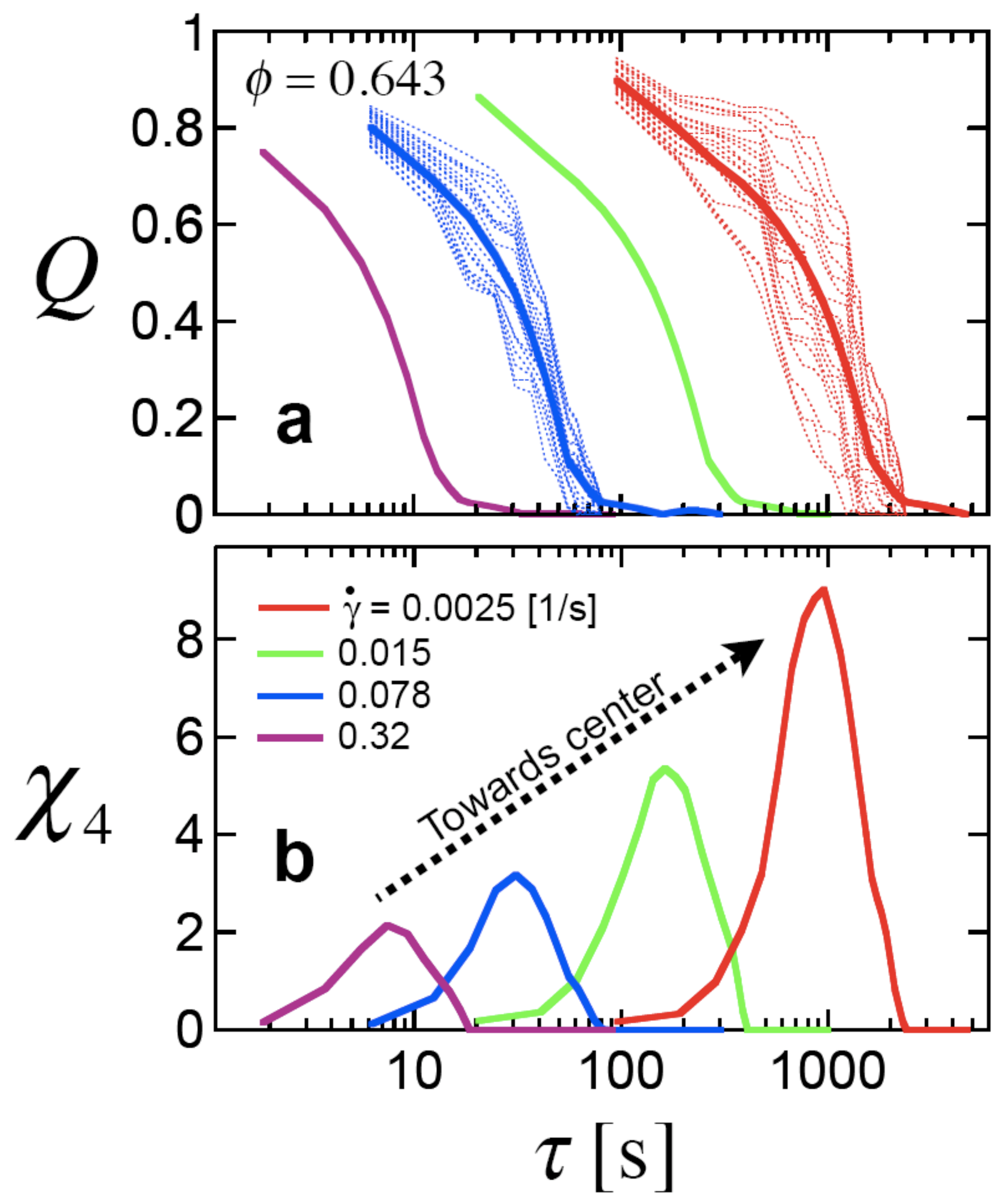}
\caption{(Color online)  (a) Overlap order parameters and (b) corresponding dynamic susceptibilities, plotted vs delay time for several strain rates as labelled.  The packing fraction is $\phi=0.643$, as in Fig.~\protect{\ref{schematic}} where strain rates are seen to be lower towards the center of the channel.  In (a) the light dashed curves represent $Q_t(\tau)$ for a selection of different start times $t$, and the heavy solid curves represent the average $Q(\tau)=\langle Q_t(\tau)\rangle$ over all $t$.}
\label{Q_x4}
\end{center}
\end{figure}

\begin{figure}
\begin{center}
\includegraphics[width=3in]{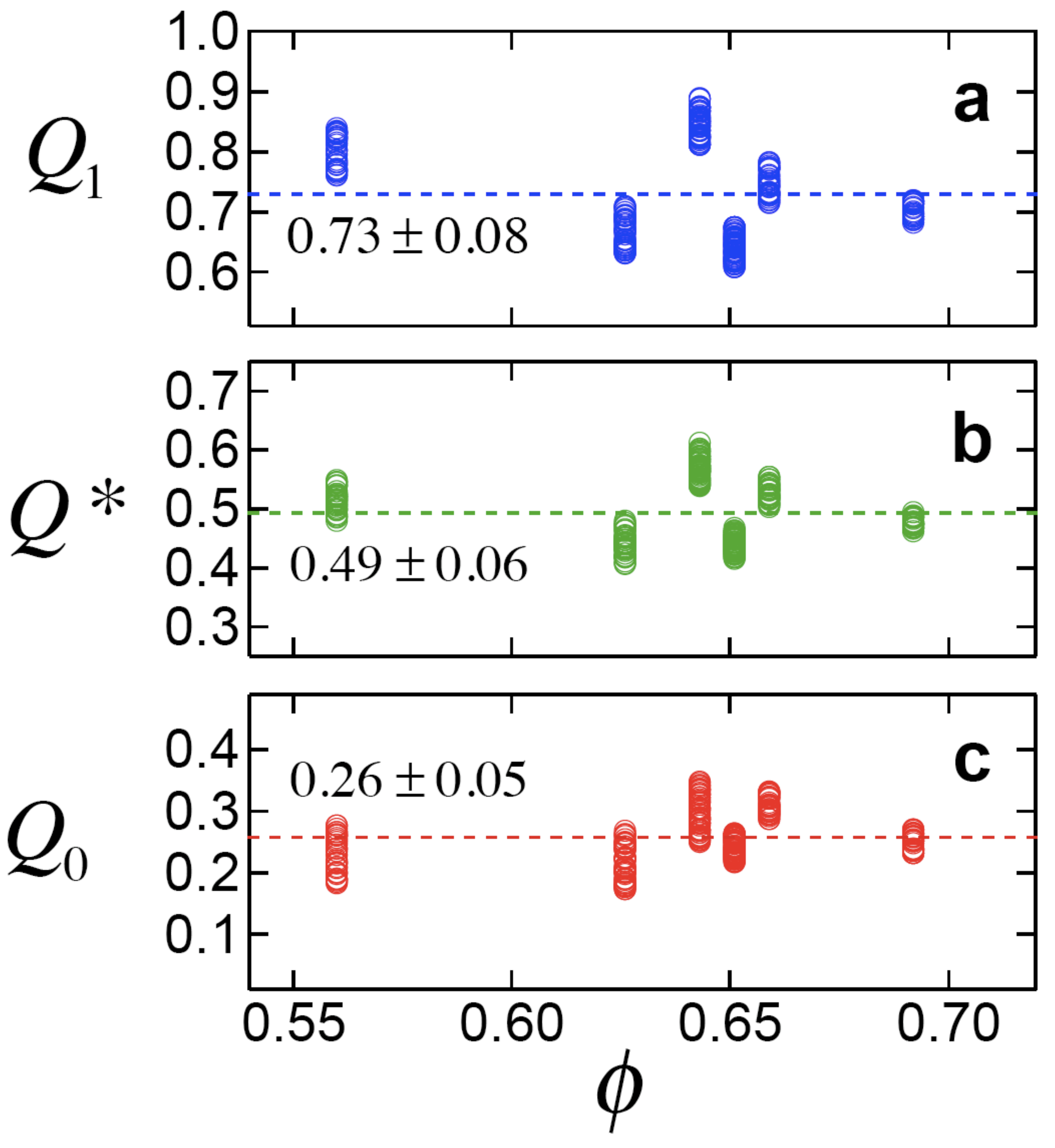}
\caption{(Color online) Averages of the overlap order parameter at time delay $\tau^*$, plotted versus volume fraction, where each data point represents a different strain rate: (a) $Q_1$ and (c) $Q_0$ are the averages for the slow and fast  regions, respectively,  whereas (b) $Q*=Q(\tau^*)$ is the average for the whole sample.  None of these quantities is found to depend on volume fraction or strain rate; their averages are indicated by the dashed horizontal lines with accompanying values.}
\label{Q}
\end{center}
\end{figure}

Spatiotemporally heterogeneous dynamics may be characterized by a four-point dynamic susceptibility, $\chi_4(\tau)$, which exhibits a peak that grows in proportion to the size of the heterogeneities.  See for example Refs.~\cite{lacevic03, LucaDHbook} for reviews.  This method begins with an ensemble-averaged self-overlap order parameter, $Q_t(\tau)$, constructed so that the contribution from each particle decays from 1 toward 0 as time increases from $t$ to $t+\tau$ and the particle moves some prescribed distance.  If all particles experience the same dynamics, then the decay of $Q_t(\tau)$ vs $\tau$ will be independent of $t$.  But if the dynamics are heterogeneous, then the decay of $Q_t(\tau)$ will be faster or slower than the time average $Q(\tau) \equiv \langle Q_t(\tau)\rangle$ according to the number of fast mobile regions that happen to exist at a particular instant.  Since the mobile regions are independent, this is governed by counting statistics and the variance
\begin{equation}
	\chi_4(\tau) \equiv N \left[ \langle Q_t^2(\tau) \rangle - \langle Q_t(\tau) \rangle^2 \right]
\label{x4}
\end{equation}
is independent of the number $N$ of particles in the system.  The average number $n^*$ of particles in a fast mobile region has been explicitly computed in Ref.~\cite{abate07} as
\begin{equation}
n^*=\frac{\chi_4^*}{(Q_1-Q_0)(Q_1-Q^*)}
\label{nstar}
\end{equation}
where $\chi_4^*$ is the peak height $\chi_4(\tau^*)$, $Q^*$ is the value of $Q(\tau^*)$, and  $Q_1$ and $Q_0$ are respectively the average values for the slow ($Q>Q^*$) and fast ($Q<Q^*$) regions.  The same results for $n^*$ were found for three very different choices for overlap order parameters, whose associated susceptibilities had different peak heights and peak times: step function, persistent area, persistent bond \cite{abate07}.  Therefore, as long as the prescription of Eq.~(\ref{nstar}) is followed, the choice of overlap order parameter is not crucial.

Since the video data (eg as in Fig.~\ref{schematic}) have insufficient resolution to track individual particle positions, we adopt an overlap order parameter similar to that introduced in Ref.~\cite{katsuragi10} based on image correlations.  In particular, we divide the video images into 50 narrow strips of constant speed and strain rate, 7~pixels $\approx 0.5~\mu$m wide, containing $N\approx 200$ particles.  For each of these strips we compute
\begin{equation}\label{qcorr}
Q_t(\tau) \equiv \frac{\langle I_i(t)I _{i+{\rm d}i}(t+\tau)\rangle-\langle I_i(t)\rangle^2}{\langle I_i(t)^2\rangle-\langle I_i(t)\rangle^2}
\end{equation}
where $\langle \cdots \rangle$ is the ensemble average over all pixels $i$ running along the strip,  where ${\rm d}i=v\tau/l$, and where $l$ is the pixel size.  Prior to this, the speed $v$ of the strip was found by varying ${\rm d}i$ at fixed $\tau$, and averaging over $t$, to maximize the cross-correlation as in the usual PIV method.  Note that the length scale probed by the associated four-point susceptibility is set by the particle-size grayscale variations in the video images.   Therefore the time $\tau^*$ at which $\chi_4(\tau)$ reaches its peak is a characteristic relaxation time needed for particle-scale relative displacements.  For illustration, example results for $Q_t(\tau)$ vs $\tau$ are shown in Fig.~\ref{Q_x4}a for a strip corresponding to packing fraction $\phi=0.643$ and strain rate $\dot\gamma=0.0025~{\rm s}^{-1}$.  This is close to jamming, and indeed the decay is quite variable.  Multiplying the variance by $N$ gives the susceptibility shown in Fig.~\ref{Q_x4}b.  This exhibits a peak at delay time $\tau^*\approx 1000$~s of height $\chi_4^*\approx 9$, when the average overlap order parameter is $Q^*\approx0.5$.  For a second example strip with a higher strain rate, $\dot\gamma=0.078~{\rm s}^{-1}$, the dynamics are more homogeneous as seen in Fig.~\ref{Q_x4} by the tighter spread of $Q_t(\tau)$ and the smaller susceptibility.

To deduce the number $n^*$ of particles in a fast mobile region from the peak height $\chi_4^*$ using Eq.~(\ref{nstar}), we must first find the three different averages of the overlap order parameter.  These are shown in Fig.~\ref{Q} as a function of volume fraction, where each point corresponds to a different strip and hence to a different strain rate.  Since there is no evident variation with volume fraction, or strain rate, we simply compute a total average over all conditions.  The average overlap order parameter at the time $\tau^*$ when $\chi_4$ peaks is found to be $Q^* \equiv Q(\tau^*) = 0.49\pm0.06$.  The average for the slow regions,  where $Q_t(\tau^*)>Q^*$, is  $Q_1=0.73\pm0.08$.  The average for the fast regions, where $Q_t(\tau^*)<Q^*$, is $Q_0=0.26\pm0.05$.  These three averages are indicated by dashed horizontal lines in Fig.~\ref{Q}.

Since the overlap order parameter is measured in a long strip, about half a particle wide and 200 particles long, the field of view may not contain the entirety of any of the fast moving regions.  However if the heterogeneities are 1-dimensional string-like swirls, as expected for a quiescent system, then they will cut the field of view a number of times in proportion to their length.  Therefore, the true number of particles involved would be a constant multiplicative factor larger than the $n^*$ value we deduce by the above prescription.  This is borne out by the good comparison of the size of heterogeneities in a monolayer of air-fluidized beads, analyzed across the whole sample or in strips \cite{LynnPressurePreprint}.  But it could also be that in channel flow the heterogenieties are linear chains or sheets of particles aligned with the velocity, in which the true number of particles involved would still be proportional to the $n^*$ value we deduce.

 
\section{Results}

\begin{figure}
\includegraphics[width=3in]{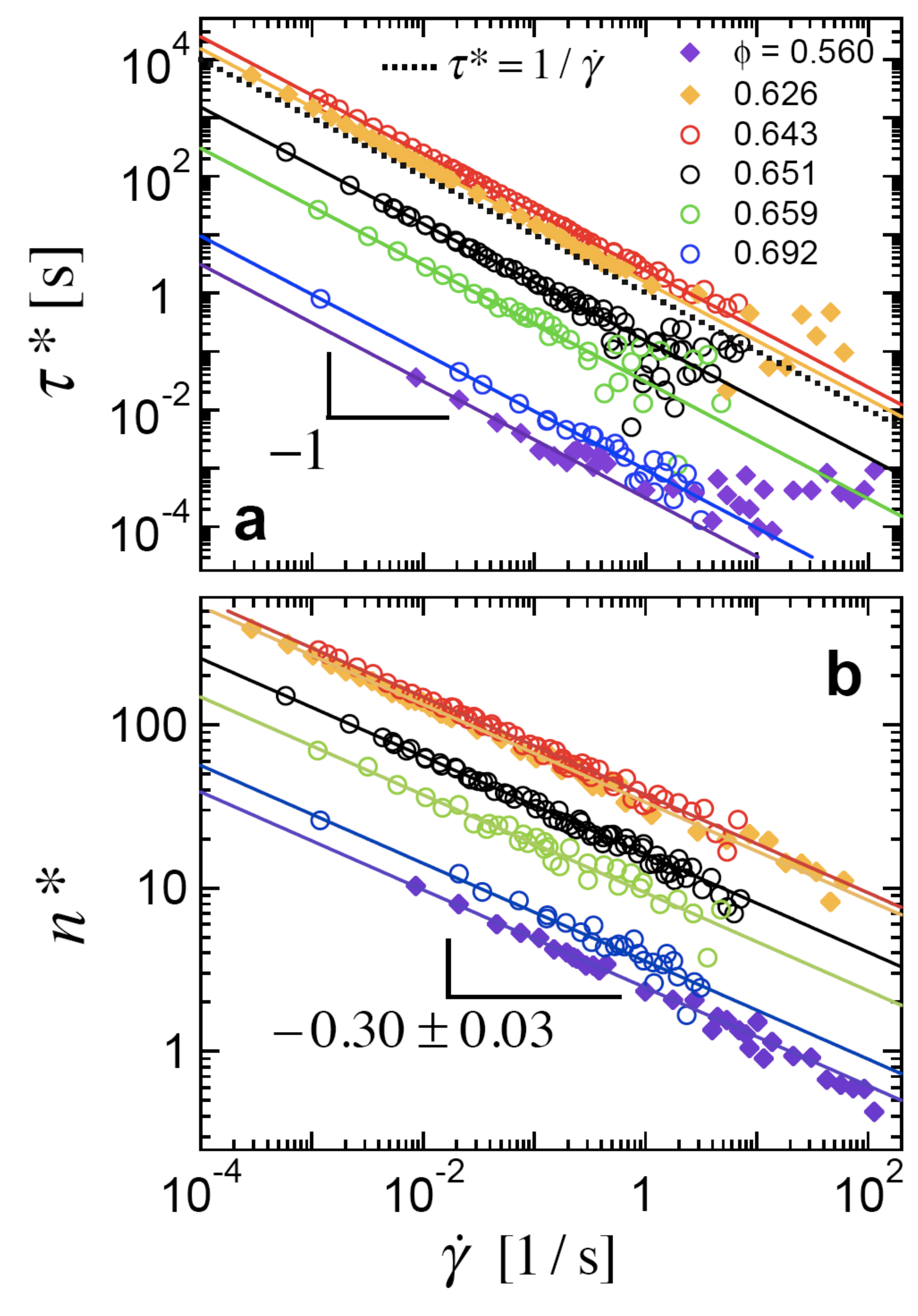}
\caption{(Color online)  (a) Relaxation time and (b) number of particles in a fast-moving heterogeneity, plotted versus strain rate, for several volume fractions $\phi$ as labelled.  The solid lines are fits to a power of $\dot\gamma$, with exponents of $-1$ in (a) and of $-0.3$ in (b).}
\label{taustar_nstar}
\end{figure}

\begin{figure}
\includegraphics[width=3in]{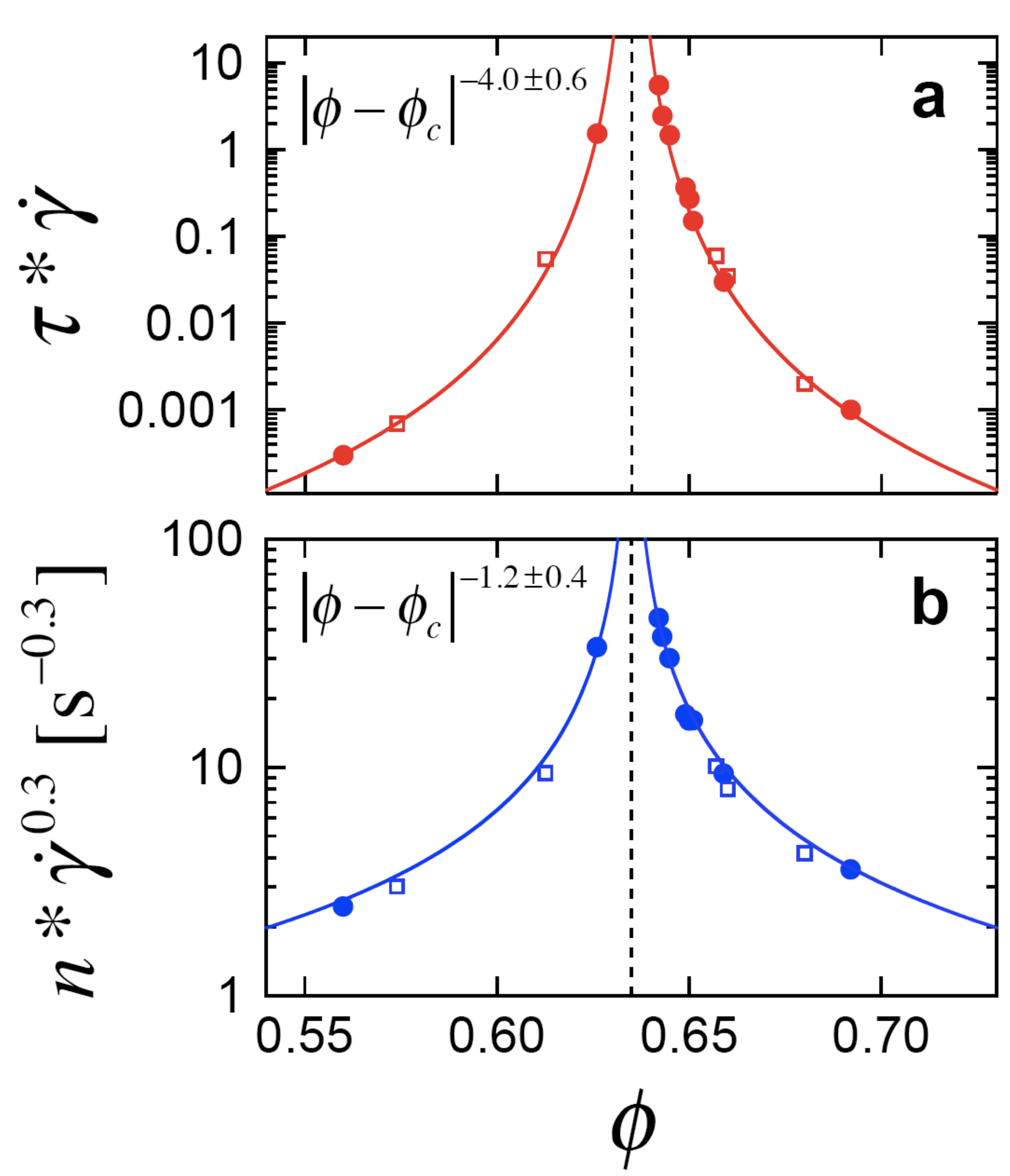}
\caption{(Color online) Coefficients of the fits in Fig.~\protect{\ref{taustar_nstar}} to (a) $\tau^*\propto1/\dot\gamma$ and (b) $n^*\propto1/\dot\gamma^{0.3}$ vs volume fraction, shown as closed symbols, for the  $\sim 1~\mu$m diameter particles.  The open symbols are for the smaller $\sim 0.6~\mu$m diameter particles.  The solid curves are fits to a power of $\Delta\phi \equiv |\phi-\phi_c|$, where $\phi_c=0.635$ is the random close packing fraction, with exponents of $-4$ in (a) and of $-1.2$ in (b). } 
\label{fitparams}
\end{figure}

\begin{figure}
\includegraphics[width=3in]{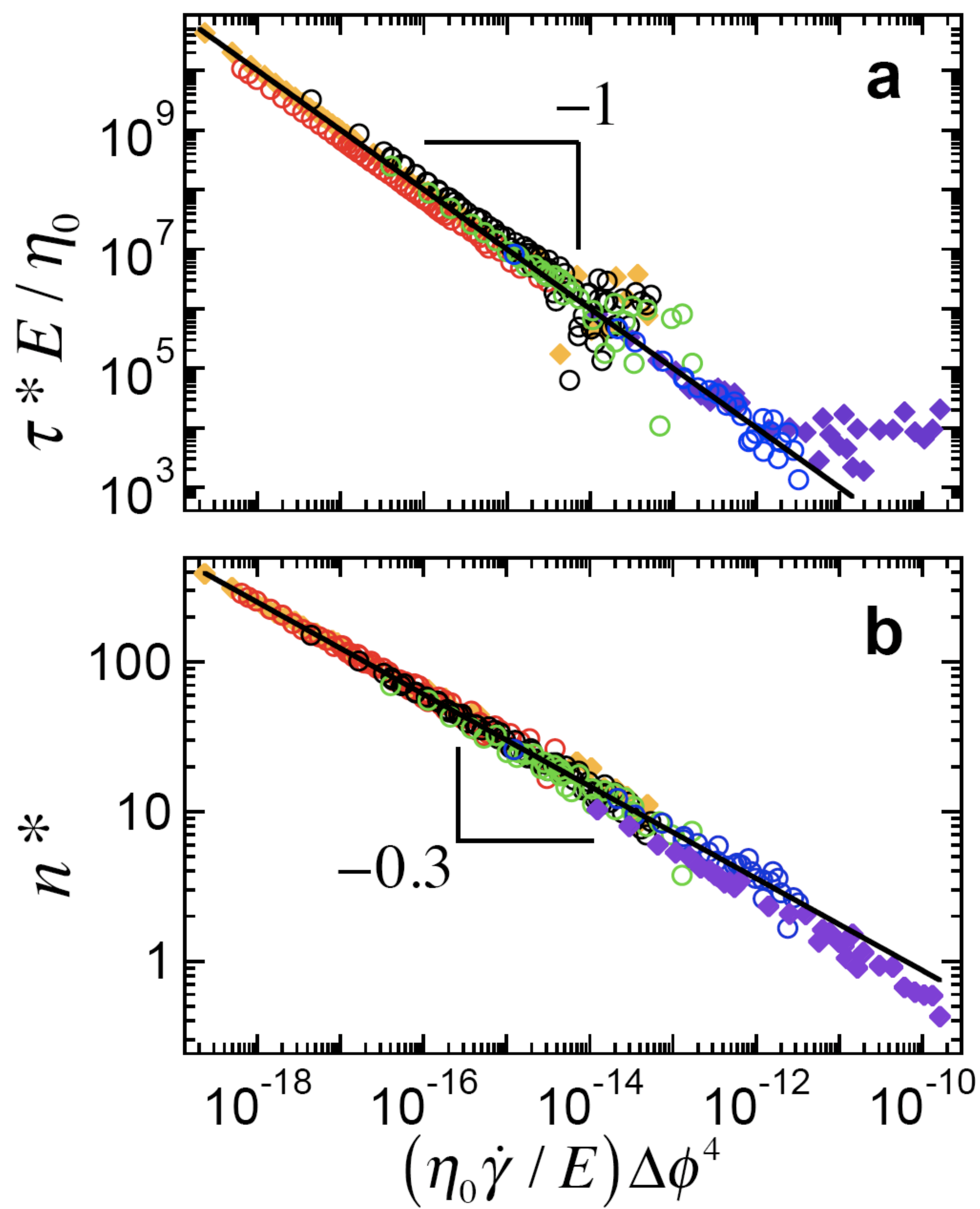}
\caption{(Color online) Scaling collapse of (a) dimensionless relaxation time and (b) number of particles in a fast-moving heterogeneity versus dimensionless strain rate times $\Delta\phi \equiv |\phi-\phi_c|$ to the fourth.  Here $\eta_0$ is the viscosity of water and $E$ is the Young elastic modulus of the particulate material.  The symbol types are the same as in Fig.~\protect{\ref{taustar_nstar}}.} 
\label{collapse}
\end{figure}

Data for the relaxation time $\tau^*$ and the number $n^*$ of particles in a fast-moving region are plotted vs strain rate in Figs.~\ref{taustar_nstar}a-b, respectively, for the $\sim 1~\mu$m diameter colloidal microgel particles.  There, each data set corresponds to a given packing fraction as labelled, and each data point corresponds to a different strip and hence to a different strain rate.  As the strain rate decreases and jamming is approached, both $\tau^*$ and $n^*$ grow as powers of the strain rate.  For the time scale, the power law is consistent with $\tau^* \propto 1/\dot\gamma$ as shown by the solid lines.  This is the simplest dimensionally-correct possibility.  For the size scale, all power law fits are consistent with $n^* \propto 1/\dot\gamma^{0.30\pm0.03}$.  This exponent agrees with recent observations of a value near $1/3$ in dry granular systems, including experiments on heap flow \cite{katsuragi10}, simulations of uniform shear \cite{Hatano0804}, and simulations of flow down an incline \cite{Staron10}.  In these works, shear occurs at essentially fixed packing fraction near $\phi_c$.  A value in the range $0.2-0.3$ was also reported for a Lennard-Jones system \cite{Tsamados10}.  So our observations considerably reinforce and extend all these results, not just to overdamped systems but also to packing fractions away from $\phi_c$ both above and below.  Nonetheless, the value of approximately $1/3$ has yet to be explained.

The relaxation time and number of particles in a fast-moving region also depend on packing fraction, as well as strain rate.  This can be seen already in Figs.~\ref{taustar_nstar}a-b, where the data sets shift up and then down as $\phi$ goes from below to above $\phi_c$.  This is displayed more clearly in Fig.~\ref{fitparams}a-b, where the coefficients $\tau^*\dot\gamma$ and $n^*\dot\gamma^{0.3}$ of the power law fits in the previous figure are plotted vs $\phi$.  The results grow without apparent bound as $\phi_c$ is approached from either side.  Data for the smaller particles are also included, and display the same behavior.  These divergences are well described by fits to power laws in $\Delta \phi = |\phi-\phi_c|$, as shown, giving $\tau^*\propto 1/\Delta\phi^{4.0\pm0.6}$ and $n^* \propto 1/\Delta\phi^{1.2\pm0.4}$.  

Altogether we thus find that the relaxation time and the size of fast-moving heterogeneities grow as the jamming transition is approached as functions of strain rate and packing fraction as
\begin{eqnarray}
	\tau^* &\propto& (\dot\gamma \Delta \phi^4)^{-1}, \label{timescaling} \\
	n^* &\propto& (\dot\gamma \Delta \phi^4)^{-0.3}. \label{sizescaling}
\end{eqnarray}
Notice that the combination $(\dot\gamma \Delta \phi^4)$ controls the behavior in both cases.  Hence there is more sensitivity to variation of $\Delta\phi$ than to variation of strain rate.  This is qualitatively consistent with numerical results for a driven kinetically constrained jamming model \cite{YairEPL10}.  To emphasize this feature, we plot all $\tau^*$ and $n^*$ results versus $(\dot\gamma \Delta\phi^4)$ in Figs.~\ref{collapse}a-b, where $\dot\gamma$ is rendered dimensionless by the intrinsic time scale set by the ratio $\eta_0/E$ of liquid viscosity to particle modulus.  Note that this collapses the data onto power laws with exponents $-1$ and $-0.3$, respectively. Thus, there are only three exponents to explain rather than four.  As discussed already, the $-1$ makes dimensional sense and the $-0.3$ extends prior observations but is not understood.   The remaining exponent, $4$, is reminiscent of the exponent  $\Gamma=4$ in the timescale $\eta_0 / (E \Delta\phi^\Gamma)$  used in Olsson-Teitel scaling plots of the shear rheology \cite{Nordstrom2010}.

We note that some simulation results are inconsistent with Eqs.~(\ref{timescaling}-\ref{sizescaling}).  In Ref.~\cite{Tsamados10} a Lennard-Jones systems shows the same size scaling, $\chi_4^* \sim 1/\dot\gamma^{0.3}$, but the relaxation time scales quite differently, $\tau^* \sim 1/\dot\gamma^{0.5}$.  In Ref.~\cite{HeussingerEPL10} harmonically-repulsive particles under quasistatic shear show $\chi_4^*\sim1/\Delta\phi^{1.8}$; the system at nonzero strain rates \cite{HeussingerSM10} shows $\chi_4^* \sim 1/\dot\gamma^{0.5-0.7}$, though the dependence on strain rate may not be a power law.


\section{Discussion}

\begin{figure}
\includegraphics[width=3in]{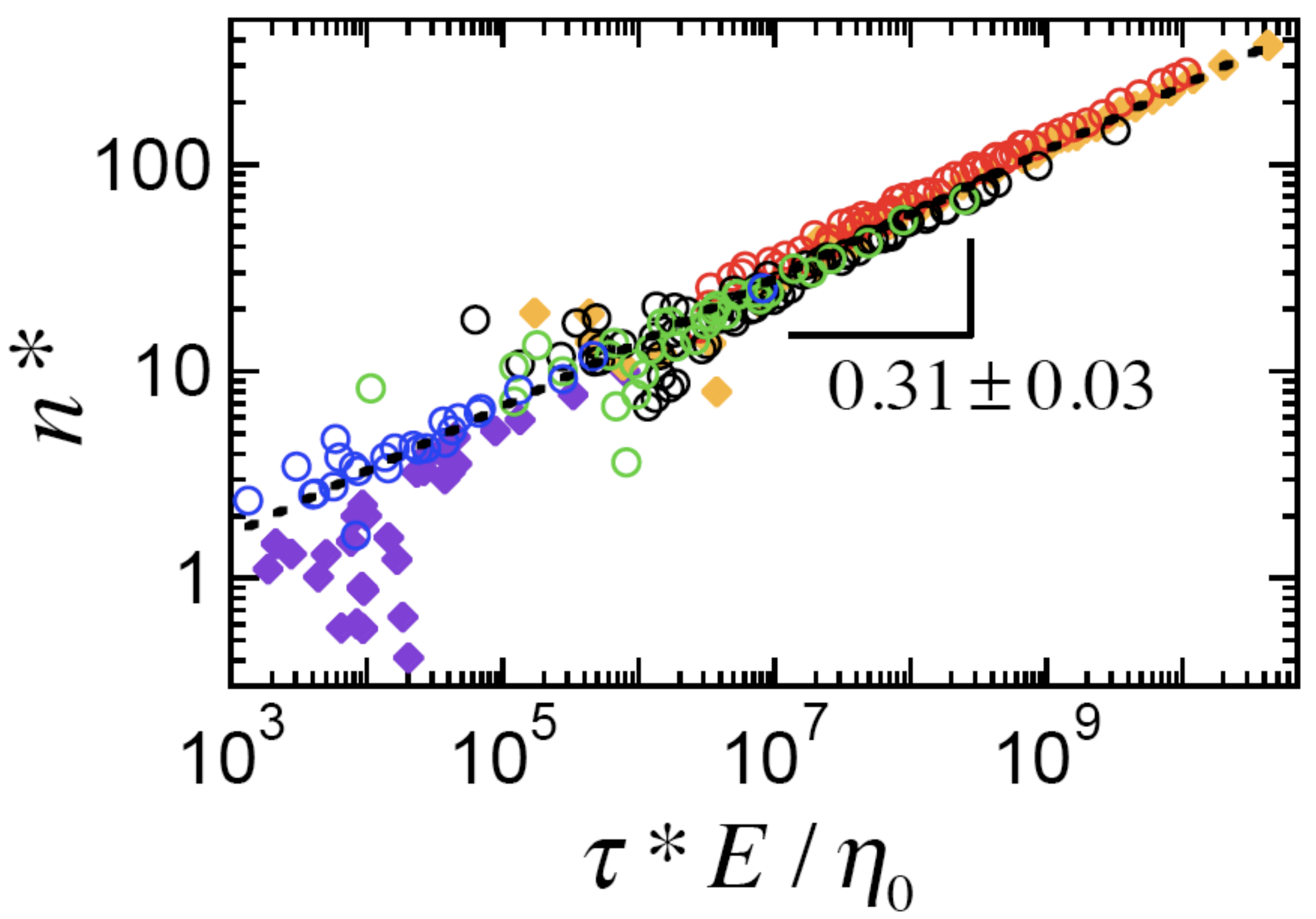}
\caption{(Color online) Number $n^*$ of particles in a fast-moving heterogeneity vs relaxation time $\tau^*$, made dimensionless by the Young modulus $E$ of the particulate material and the viscosity $\eta_0$ of the suspending fluid (water).  The dashed black line is a power-law fit with exponent $\lambda=0.31\pm0.03$ as labeled.  The symbol types are the same as in Fig.~\protect{\ref{taustar_nstar}} 
} 
\label{nstartaustar}
\end{figure}

The observations made here, summarized by Eqs.~(\ref{timescaling}-\ref{sizescaling}), combine to give the size of heterogeneities as a power-law of the relaxation time:
\begin{equation}
	n^*\propto (\tau^*)^{\lambda}.
\label{nvstau}
\end{equation}
For emphasis, we plot $n^*$ data vs $\tau^* E/\eta_0$ in Fig.~\ref{nstartaustar} on logarithmic axes and observe that indeed the data collapse to a straight line.  Fitting to a power-law gives the exponent and uncertainty as
\begin{equation}
	\lambda=0.31\pm0.03.
\label{lambda}
\end{equation}
Note that the dynamic range is more than two decades in size and seven in relaxation time, which is sufficient to rule out the possibilities of a logarithmic or exponential connection between $n^*$ and $\tau^*$.   The power law form is consistent with mode-coupling theory, and the exponent is only slightly smaller than the expectation $1/\gamma$ reported in Refs.~\cite{KobPRL94, WhitelamPRL04}.

One advantage of plotting $n^*$ and $\tau^*$ parametrically versus one another, rather than versus the control parameters, is that it allows comparison with other systems where there is no shear or where the control parameter is something other than strain rate.  For example, $\lambda=1/2$ is reported for the driven kinetically constrained model mentioned above \cite{YairEPL10}, while a logarithmic connection better accounts for the simulations of Brownian harmonically-repulsive particles \cite{HaxtonLiuEPL10}.  In terms of experiment, comparison is possible for only a few experiments of which we are aware.  For hard spheres, Ref.~\cite{BrambillaPRL09} found that $\tau^*$ grows faster than a power law, and that $\chi_4^*$ grows slower than a power law, as $\phi$ approaches $\phi_c$ from below.  It is stated that $n^*$ grows logarithmically with $\tau^*$, which disagrees with our results.  For soft NIPA microgel particles similar to those studied here, Ref.~\cite{sessoms09} found that both $\tau^*$ and $\chi_4^*$ grow with increasing packing fraction; no functional form was proposed or tested.  We digitized their data and plot parametrically, rather than vs packing fraction, and find power law behavior of the form Eq.~(\ref{nvstau}) with exponent $\lambda=0.34\pm0.16$.  This is consistent with our findings.  For a monolayer of large spherical grains fluidized by a steady upflow of air, Ref.~\cite{AbatePRL08} found that there is a meaningful effective temperature $T_{\rm eff}$, and that size and time scales are consistent with $n^*\sim 1/{T_{\rm eff}}^{0.7\pm0.2}$ and $\tau^* \sim 1/{T_{\rm eff}}^{2\pm0.5}$, respectively.  These combine to give a power law relationship, Eq.~(\ref{nvstau}), with exponent $\lambda=0.35\pm0.15$ that agrees with the results here in Fig.~\ref{nstartaustar}.   Further experiments on the fluidized grains, where the sample is tilted and where the analysis is carried out by dividing the sample into a series of strips each at a different pressure, also appear to agree \cite{LynnPressurePreprint}.  For steady gravity-driven flow of grains down along a confined heap, and visualized through the sidewalls as a function of depth $z$ below the free surface, Ref.~\cite{katsuragi10} found $\tau^*\propto 1/I$ and $n^*\propto (1/I)^\lambda$ where $I=\dot\gamma d/\sqrt{gz}$ is the inertia number, $d$ is the grain diameter, $g=9.8$~m/s$^2$, and $\lambda=0.33\pm0.02$.  This system is underdamped, so the time scale is rendered dimensionless by different microscopic physics, but the exponent for the power-law connection between $n^*$ and $\tau^*$ is the same as found here.


\vfill

\section{Conclusion}

In this paper we presented a study of dynamical heterogeneities in a colloidal system that (a) is not hard spheres, (b) is compressed above as well as below $\phi_c$, and (c) is subject to shear.  Our experiments made crucial use of custom synthesized NIPA microgel particles and of a custom fabricated microfluidic channel, as well as of a novel video-based dynamical order parameter.   As jamming is approached by bringing the packing fraction difference $\Delta\phi=|\phi-\phi_c|$, or the strain rate $\dot\gamma$, to zero, we demonstrated that the time and size scales for dynamical heterogeneities both grow as powers of the combination ($\dot\gamma\Delta\phi^4$) according to Eqs.~(\ref{timescaling}-\ref{sizescaling}).  While there is precedence for the observed strain rate dependence from experiments on underdamped granular systems, the packing fraction dependence appears to be a new result.  The observed connection between the size and time scale is a power law, $n^* \propto (\tau^*)^\lambda$, consistent with mode coupling theories but not with observations for an unsheared suspension of Brownian hard sphere colloids.  It is intriguing that the exponent we find, $\lambda\approx 1/3$, agrees with precedents for unsheared soft particles \cite{sessoms09} and for two driven systems of hard grains -- one with shear \cite{katsuragi10} and one without \cite{AbatePRL08, LynnPressurePreprint}.  This suggests universality with respect to interactions, but with unsheared hard sphere colloids in a different universality class.


\begin{acknowledgments}
This work was supported by the National Science Foundation through grants MRSEC/DMR05-20020 and DMR-0704147.  We thank L. Cipelletti, A.J. Liu, and P. Yunker for helpful conversations.  We thank A. Alsayed, A. Basu, and Z. Zhang for their help in synthesizing the particles.
\end{acknowledgments}

\bibliography{References_dynamicalhet}

\end{document}